\documentclass[aps,prl,showpacs,nofootinbib, twocolumn]{revtex4}

\usepackage{amssymb}
\usepackage{amsmath}
\usepackage{bm}
\expandafter\ifx\csname package@font\endcsname\relax\else
\expandafter\expandafter
 \expandafter\usepackage
 \expandafter\expandafter
 \expandafter{\csname package@font\endcsname}%
\fi
\renewcommand{\prd}{{\it Phys. Rev. D}}
\renewcommand{\prl}{{\it Phys. Rev. Lett.}}

\begin{document}
\title{Comment on "The gravitomagnetic influence on gyroscopes and on the lunar orbit"}
\author{Sergei M. Kopeikin}
\email{kopeikins@missouri.edu}
\affiliation{Department of Physics \& Astronomy, University of
Missouri-Columbia, 65211, USA}
\begin{abstract}
Analysis of the gauge residual freedom in the relativistic theory of lunar motion demonstrates that lunar laser ranging (LLR) is not currently capable to detect gravitomagnetic effects. 
\end{abstract}
\maketitle\noindent
The {\it Letter} \cite{nortur} states that the gravitomagnetic interaction plays a part in shaping the lunar orbit readily obervable by LLR. The authors pick up a ``gravitomagnetic" term from the PPN equation of motion of massive bodies \cite{will} and prove that it correctly reproduces the Lense-Thirring precession of the GP-B gyroscope. The paper \cite{nortur} argues that the very same term in the equations of motion of the Moon, derived in the solar-system barycentric (SSB) frame, perturbs the lunar orbit with a radial amplitude $\simeq 6$ meters that was observed. Below we explain that the gauge freedom of the three-body problem supresses the gravitomagnetic (and any other) effects in the lunar motion, that depend on the Earth's velocity ${\bm V}$ around the Sun, to the level $\le 1$ millimeter. It makes LLR insensitive to the gravitomagnetic interaction.

Paper \cite{nortur} uses the SSB frame to derive the equation of motion of the Moon relative to the Earth. It formally includes the gravitomagnetic pertrubation in the following form \cite{nortur,will,note1}
\begin{equation}
\label{1}
{\bm a}_{\rm GM}={\bm a}+\frac{\gamma-1}{2}{\bm a}\;,
\end{equation}
where $\gamma$ parameterizes a deviation from general relativity, 
\begin{equation}
\label{2}
{\bm a}=\frac{4GM}{c^2r^2}\left[\hat{\bm r}\left(V^2+{\bm V}\cdot{\bm u}\right)-{\bm V}\left({\bm V}\cdot\hat{\bm r}+{\bm u}\cdot\hat{\bm r}\right)\right]\;,
\end{equation}
$M$ is mass of the Earth, $r$ is radius of the lunar orbit, $\hat{\bm r}$ is the unit vector from the Earth to the Moon, ${\bm V}$ is the Earth's velocity around the Sun, and ${\bm u}$ is the Moon's velocity around the Earth. The SSB frame referred to the geocenter {\it is not} in a free fall about the Sun, and {\it does not} make a local inertial frame, as it is obtained in \cite{nortur,wtb} by means of a Newtonian-like translation from the solar-system barycenter to the geocenter.

Thus, perturbations in Eqs. (\ref{1})-(\ref{2}) can not be interpreted as physically observable and, in fact, represent a spurious gauge-dependent effect that is cancelled by transformation to the local-inertial frame of the geocenter. The gauge freedom of the lunar equations of motion must be analyzed to eliminate all gauge-dependent, non-observable terms. The analysis of the gauge freedom in the three body-problem had been done in \cite{bk,dsx}. It proves that all non-tidal and ${\bm V}$-dependent terms, including the first term in the right side of Eq. (\ref{1}), are pure coordinate effects that disappear from the lunar equations of motion after transformation to the geocentric, locally-inertial frame. This is because the Lorentz invariance and the principle of equivalence reduce the relativistic equation of motion of the Moon to the covariant equation of the geodesic deviation between the Moon's and the Earth's world lines where gravitomagnetic effects appear only as tidal relativistic forces with amplitude much smaller than 1 millimeter \cite{bk,dsx}. The covariant nature of gravity tells us \cite{will} that if some effect is not present in the local frame of observer, it can not be observed in any other frame. This means that besides physically-observable terms, the LLR model \cite{nortur,wtb} also operates with terms having gauge-dependent origin, which mathematically nullify each other in the SSB frame. The mutually-annihilating terms enter different parts of the LLR model with opposite signs \cite{bk,dsx} but, if taken separately, can be erroneously interpreted as really observable \cite{nortur}. 

General relativity indicates that the SSB lunar equations of motion may admit the observable gravitomagnetic acceleration only in the form of the second term in the right side of Eq. (\ref{1}) that is proportional to $\gamma-1$. Radio experiments set a limit on $\gamma-1 \le 10^{-3}$ \cite{will} that yields $|{\bm a}_{\rm GM}|\le 1$ millimeter. The current half-centimeter accuracy of LLR is insufficient to measure such negligible effect. We conclude that LLR is currently insensitive to the gravitomagnetism and, hence, can not compete with the GP-B experiment.     


\begin{thebibliography}{}
\bibitem{nortur}T.~W.  Murphy,  Jr., 
K. Nordtvedt, and S.G. Turyshev, \prl, {\bf 98}, 071102 (2007)
\bibitem{will}C.~M. Will,  "The Confrontation between General Relativity and Experiment", Living Rev. Relativity {\bf 9}, 3 (2006). 
\bibitem{note1} Bold letters denote spatial vectors. Euclidean dot product of two vectors is denoted as ${\bm a}\cdot{\bm b}$.
\bibitem{wtb} J.G. Williams, S.G. Turyshev, and D.H. Boggs, arXiv:gr-qc/0507083
\bibitem{bk} V.A. Brumberg, and S.M. Kopeikin, {\it Nuovo Cim. B}, {\bf 103}, 63 (1989) 
\bibitem{dsx} T. Damour, M. Soffel, and C. Xu, {\prd}, {\bf 49}, 618 (1994)
\end{thebibliography}
\end{document}